\documentclass[aps,prl,twocolumn,groupedaddress]{revtex4}
\usepackage{amsmath}
\usepackage{amssymb}

\input{tcilatex}

\begin{document}

\title{Stiff spinning torus of electromagnetic two-body motion}
\author{Jayme De Luca}
\email[author's email address: \ ]{deluca@df.ufscar.br}
\affiliation{Universidade Federal de S\~{a}o Carlos, \\
Departamento de F\'{\i}sica\\
Rodovia Washington Luis, km 235\\
Caixa Postal 676, S\~{a}o Carlos, S\~{a}o Paulo 13565-905}
\date{\today }

\begin{abstract}
We study an orbit of the electromagnetic two-body problem that involves a
fast (stiff) spinning motion about a circular orbit. We give a multiscale
method of solution that solves for the fast timescale first. The solvability
condition of the asymptotic expansion demands resonances for the fast
dynamics. The stiff resonant tori are found precisely in the atomic
magnitude and agree with many features of the Bohr atom in quantitative and
qualitative detail; We calculate every first emission line of the first $13$
observable spectroscopic series of hydrogen within a few percent deviation.
The resonant orbits have angular momenta that are approximate multiples of
Planck's constant and the emitted frequencies are given by a difference of
two linear eigenvalues. This Lorentz-invariant two-body dynamics exhibts the
phenomenon of resonant dissipation, i.e., the metastable dynamics radiates
the center-of-mass energy while the particles perform fast spinning
oscillations of small amplitude about a circular orbit, a collective
radiative recoil.
\end{abstract}

\maketitle

We study a two-body orbit of classical electrodynamics involving a fast
(stiff) spinning toroidal motion about a circular orbit. The nontrivial
stiff dynamics is part of the full two-body motion with delay and it
disappears in the Coulombian limit.\ Our equations of motion are prescribed
by Dirac's electrodynamics of \emph{point} charges with retarded-only fields 
\cite{Dirac}, a fundamental physical theory. The existence of the fast
delay-dynamics demands a multiscale asymptotic solution that solves for the
fast timescale first. The multiscale method requires a resonance between two
fast frequencies because of Freedholm's alternative \cite{PRA}, providing
the natural coupling between the fast and the slow timescales. The resonant
orbits turn out to be precisely in the atomic magnitude; These resonant
orbits are labelled by an integer and have angular momenta that are
multiples of a basic angular momentum, that agrees well with Planck's
constant. The emitted frequencies agree with the sharp lines of quantum
electrodynamics (QED) for the circular lines of hydrogen with a few percent
average deviation. The angular momentum of the stiff spinning motion is of
the order of the orbital angular momentum, again in qualitative agreement
with QED.

Dirac's 1938 work \cite{Dirac} on the electrodynamics of \emph{point}
charges gave delay equations that were seldom studied. Among the few
dynamics investigated, an early result of Eliezer \cite{Eliezer, Andrea,
Massimo} revealed a surprising result (henceforth called Eliezer's theorem);
An electron moving in the Coulomb field of an infinitely massive proton can
never fall into the proton by radiating energy. The result was generalized
to tridimensional motions with self-interaction in a Coulomb field \cite%
{Andrea, Massimo}, finding that only scattering states are possible. Since
our model has the dynamics of Eliezer's theorem as the infinite-mass limit,
a finite mass for the proton is essential for a physically meaningful
dynamics; If the proton has a finite mass, there is no inertial frame where
it rests at all times, and this in turn causes delay because of the finite
speed of light. A finite mass for the proton is what brings delay into the
electromagnetic two-body dynamics, with its associated fast dynamics. We
shall describe the two-body motion in terms of the familiar \emph{%
center-of-mass coordinates} and \emph{coordinates of relative separation},
defined as the familiar coordinate-transformation that maps the two-body
Kepler problem onto the one-body problem with a reduced mass plus a
free-moving center of mass. We stress that in the present relativistic
motion the Cartesian center-of-mass vector is not ignorable, and it
represents three extra coupled degrees-of-freedom. We introduce the concept
of resonant dissipation to generalize the concept of non-ionizing orbits of
Ref.\cite{Darwin} . Resonant dissipation is the condition that both
particles recoil together, i.e., the center-of-mass vector decelerates,
while the particles perform the almost-circular orbit illustrated in Fig. 1,
despite of the energy losses of the metastable center-of-mass dynamics.

Here we use a circular orbit as the unperturbed solution of our perturbation
scheme. This circular orbit is a solution of the low-velocity
Coulombian-limit obtained by ignoring the delay and its associated fast
dynamics. The quantities for this uniform circular motion are given now \cite%
{dissipaFokker}; The electron has mass $m_{1}$ and travels a circle of
radius $r_{1}$ while the proton has mass $m_{2}$ and travels a circle of
radius $r_{2}$. We henceforth use units where the electronic charge is $e=-1$
and the speed of light is $c=1$. The circular orbit is defined by the
retardation angle $\theta $ that one particle turns while the light
emanating from the other particle reaches it (one light-cone distance away).
The angular momentum $l_{z}$ of the circular orbit is given to first order
by $l_{z}^{-1}=\theta $, and for atomic orbits $\theta $ is a small
parameter of the order of the fine structure constant $\alpha =1/137.036$.
For small values of $\theta $ the frequency of angular rotation $\Omega $
and the particle distance in light-cone $r_{b}$ are given, to the leading
order in $\theta $, by Kepler's law%
\begin{eqnarray}
\Omega &=&\mu \theta ^{3}+...  \label{Kepler} \\
r_{b} &=&\frac{1}{\mu \theta ^{2}}+...  \notag
\end{eqnarray}%
where $\mu \equiv m_{1}m_{2}/(m_{1}+m_{2})$ is the reduced mass. The
circular orbit is not an exact solution of the full two-body equations of
Dirac's theory\cite{Dirac} with a finite protonic mass and self-interaction;
Substituting the circular orbit into the equations of motion of \cite{Dirac}%
, we find for the electronic motion an offending force opposite to the
electronic velocity of magnitude%
\begin{equation}
r_{b}^{2}F_{1}=r_{b}^{2}(\frac{2}{3}\dddot{x}_{1})\simeq -\frac{2}{3}\theta
^{3},  \label{offending}
\end{equation}%
where we scaled the offending force by the magnitude of the attractive force
between the particles, a scaling factor of $1/r_{b}^{2}$ . Along the
trajectory of the proton, the delayed Li\`{e}nard-Wiechert attraction of the
electron is the main offending force against the velocity, a force of
exactly the same magnitude of Eq. (\ref{offending}) as calculated with the
Page expansion of Ref.\cite{PRL}. These offending forces cause dissipation
along the unperturbed circular orbit. This radiative instability in the slow
timescale is an incomplete picture of the dynamics though; it disregards the
fast stiff motion that is present because of the delay \cite{dissipaFokker}.
Precisely because of the stiff delay dynamics, at very small deviations from
circularity the nonlinear terms of the self-interaction force correct the
leading term of Eq. (\ref{offending}), enough to balance the offending force
of Eq. (\ref{offending}). In the following we explain in physical terms how
the stiff torus of Fig.1 is formed.

We start by substituting the circular orbit plus a perturbation into the
equations of motion of \cite{Dirac}. For the stability analysis we transform
the Cartesian coordinates of the particles to gyroscopic coordinates

\begin{eqnarray}
x_{k}+iy_{k} &\equiv &r_{b}\exp (i\Omega t)[d_{k}+\xi _{k}^{\ast }],  \notag
\\
x_{k}-iy_{k} &\equiv &r_{b}\exp (-i\Omega t)[d_{k}+\xi _{k}],  \notag \\
z_{k} &\equiv &r_{b}(Z_{k}+Z_{k}^{\ast })  \label{gyroscopic}
\end{eqnarray}%
where $Z_{k}$ and $\xi _{k}$ are complex numbers defining the perturbations
to the circular orbit. In Eq. (\ref{gyroscopic}) the coordinates of the
electron are defined by $k=1$ and $d_{1}\simeq (m2+m1)/m2$ \ while the
coordinates of the proton are defined by $k=2$ and $d_{2}\simeq -(m2+m1)/m1$ 
\cite{dissipaFokker}. Notice that the perturbations of Eq. (\ref{gyroscopic}%
) are scaled by the light-cone distance $r_{b}$. The linear stability
analysis proceeds like in Refs. \cite{StaruszkiewiczPole, dissipaFokker}, by
substituting Eq. (\ref{gyroscopic}) into the equations of motion and
expanding the equations of motion to linear order.\ Analogously to Ref. \cite%
{dissipaFokker}, the tangent dynamics perpendicular to the plane (the
z-direction) is decoupled from the tangent dynamics along the orbital plane
to linear order. As found in Ref \cite{dissipaFokker}, the linearized
equations of motion have normal-mode solutions with every coordinate
proportional to $\exp (\lambda \Omega t/\theta )$. We shall see that $%
|\lambda |$ can be of the order of $\pi $ or larger, such that $\lambda
\Omega /\theta $ is a frequency much larger than the orbital frequency $%
\Omega $ (here the stiffness enters). Again a remarkable quasi-degeneracy is
found between the planar and the perpendicular normal-mode conditions\cite%
{dissipaFokker} in the stiff-limit. If $|\lambda |$ is large enough (the
stiff-limit), both conditions reduce to

\begin{equation}
(\frac{\mu \theta ^{4}}{M})\exp (-2\lambda )=1,  \label{Istar}
\end{equation}%
where $\mu $ is the reduced mass defined below Eq. (\ref{Kepler}) and $%
M\equiv m_{1}+m_{2}$ is the total mass of the atom. For hydrogen $(\mu /M)$
is a small factor of about $(1/1824)$. Atomic orbits have $\theta $ of the
order of the fine structure constant such that the small parameter $\mu
\theta ^{4}/M\sim 10^{-13}$ multiplying the exponential function in Eq.(\ref%
{Istar}) determines that $\func{Re}(\lambda )\equiv $ $-\sigma \simeq -\ln (%
\sqrt{M/\mu \theta ^{4}})$. For the first $13$ excited states of hydrogen
this $\sigma $ is in the interval $14.0<|\sigma |<18.0$. The imaginary part
of $\lambda $ can be an arbitrarily large multiple of $\pi $, such that the
general solution to Eq. (\ref{Istar}) is 
\begin{equation}
\lambda =-\sigma +\pi qi  \label{unperastar}
\end{equation}%
where $i\equiv \sqrt{-1}$ and $q$ is an arbitrary integer. \ Notice that the
real part of $\lambda $ is always negative, such that the tangent dynamics
about circular orbits is stable in the stiff-limit with delay-only\cite%
{Dirac}, at variance with the dynamics with advance and delay of Ref. \cite%
{dissipaFokker}.

The full normal-mode condition in the stiff-limit is derived in the same way
of Ref. \cite{dissipaFokker} and it will be published elsewhere. The
perpendicular normal-mode condition is%
\begin{eqnarray}
&&(1+\frac{2}{\lambda _{z}}-\frac{1}{\lambda _{z}^{2}}-\frac{1}{\lambda
_{z}^{3}}+\frac{1}{\lambda _{z}^{4}}+...)(\frac{\mu \theta ^{4}}{M})\exp
(-2\lambda _{z})  \notag \\
&=&1-\frac{2}{3}\theta ^{2}\lambda _{z}+\frac{4\mu }{M}\theta ^{4}\lambda
_{z}^{2}+...,  \label{norZ}
\end{eqnarray}%
and the planar normal-mode condition is 
\begin{eqnarray}
&&(1+\frac{2}{\lambda _{xy}}+\frac{7}{\lambda _{xy}^{2}}+\frac{10}{\lambda
_{xy}^{3}}-\frac{5}{\lambda _{xy}^{4}}+...)(\frac{\mu \theta ^{4}}{M})\exp
(-2\lambda _{xy})  \notag \\
&=&1-\frac{2}{3}\theta ^{2}\lambda _{xy}+\frac{1}{9}\theta ^{4}\lambda
_{xy}^{2}+...,  \label{norXY}
\end{eqnarray}%
The coefficient of $\ $the term of order $1/\lambda ^{2}$ is the main\
difference between Eqs. (\ref{norZ}) and (\ref{norXY}) for $\theta $ in the
atomic range. The term on the right-hand side of Eqs. (\ref{norZ}) and (\ref%
{norXY})\ with the $2/3$ coefficient is due to the third derivative of the
self-interaction force of Eq. (\ref{offending}). The exact roots of Eqs. (%
\ref{norXY}) and (\ref{norZ}) near the limiting-root (\ref{unperastar}) are
defined respectively by 
\begin{eqnarray}
\lambda _{xy}(\theta ) &\equiv &-\sigma _{xy}+\pi qi+i\epsilon _{1},
\label{pair} \\
\lambda _{z}(\theta ) &\equiv &-\sigma _{z}+\pi qi+i\epsilon _{2},  \notag
\end{eqnarray}%
where $\epsilon _{1}(\theta )$ and $\epsilon _{2}(\theta )$ are real and
small numbers that depend on the orbit through $\theta $. The stiff torus is
formed from an initial circular orbit as follows; (i) the offending force
against the velocity makes the electron loose radius by radiating energy on
a slow timescale, and (ii) when the electron deviates enough from the
circular orbit, the stiff nonlinear terms compensate the offending force of
Eq. (\ref{offending}) and balance the real parts of Eq. (\ref{pair}), the $%
\sigma ^{\prime }s$. The motion continues to be stiff because of the large
imaginary part of the $\lambda ^{\prime }s$ of Eq. (\ref{pair}). In the
following we outline a multiscale expansion about the balanced stiff torus
to derive a resonance condition between the modes of tangent dynamics. The
circular orbit recoils in a slow timescale to compensate the momentum of the
radiated energy, such that the particle coordinates are described by a
composition of a translation mode and a stiff mode, both with slowly varying
amplitudes 
\begin{eqnarray}
\xi _{k} &=&A_{k}(T)+u_{k}(T)\exp [(\pi qi+i\epsilon _{1}^{a})\Omega
t/\theta ],  \label{multiscalesolution} \\
Z_{k} &=&B_{k}(T)+R_{k}(T)\exp [(\pi qi+i\epsilon _{2}^{a})\Omega t/\theta ],
\notag
\end{eqnarray}%
where the amplitudes $u_{k}$ and $R_{k}$ must be near the size $a$ where the
nonlinearity balances the negative real part of the linear modes. Because of
the large frequency of the stiff motion, this balancing is established at a
very small radius, as we estimate in the following.

The correction to Eq. (\ref{offending}) is provided by the nonlinear terms
of the Lorentz-Dirac force, as given in page 116 of Ref \cite{Rohrlich}.
Keeping only terms along the velocity, the self-interaction force along the
electronic velocity is%
\begin{equation}
r_{b}^{2}F_{1}=\frac{2}{3}\gamma _{1}^{3}r_{b}^{2}[\dddot{x}_{1}+\gamma
_{1}^{2}(v_{1}\cdot \dddot{x}_{1})v_{1}+...],  \label{nondissi1}
\end{equation}%
where we again multiplied by the convenient scaling factor, as in Eq. (\ref%
{offending}). Using the stiff term of Eq. (\ref{multiscalesolution}) into
Eq. (\ref{gyroscopic}) to estimate the coefficient of the velocity term on
the right-hand side of Eq. (\ref{nondissi1}) yields%
\begin{equation}
r_{b}^{2}F_{1}\simeq \frac{2}{3}[-\theta ^{3}+\left( \pi q\right)
^{4}a^{2}\theta ]=0,  \label{balance}
\end{equation}%
where we used $|u_{1}|\simeq |R_{1}|$ $\simeq a$ in Eq. (\ref%
{multiscalesolution}), $a$ being the radius of the stiff torus scaled by the
light-cone radius $r_{b}$. Equation (\ref{balance}) estimates that along the
stiff torus the stiff nonlinear term balances the offending force on the
electron at the amplitude%
\begin{equation}
a=\frac{\theta }{\pi ^{2}q^{2}}.  \label{relativeradius}
\end{equation}%
This distance is a small fraction of the orbital radius and it represent a
length of 
\begin{equation}
r_{b}a=\frac{1}{\theta \pi ^{2}q^{2}},  \label{absoluteradius}
\end{equation}%
which for $\theta $ \ in the atomic scale is less than $30$ classical
electronic radia. This is the radius of the stiff torus, a very reasonable
physical length. Notice also that the velocity associated with the stiff
motion is given by%
\begin{equation}
v=\frac{\pi q}{\theta }\Omega r_{b}a=\frac{\theta }{\pi q},
\label{velostiff}
\end{equation}%
which is much lesser than the speed of light for $\theta $ in the atomic
scale.

Assuming that $|u_{1}|=|R_{1}|=a$, as estimated by Eq. (\ref{relativeradius}%
), we can use (\ref{multiscalesolution}) to evaluate the angular momentum of
the stiff torus along the orbital plane. Disregarding fast oscillatory
terms, this angular momentum along the orbital plane is%
\begin{eqnarray}
l_{x}+il_{y} &=&\mu r_{b}^{2}a^{2}\frac{\pi q\Omega }{\theta }b_{1}\exp
[i(\epsilon _{2}^{a}-\epsilon _{1}^{a}+\theta )\Omega t/\theta ]
\label{angularmomentum} \\
&=&\frac{\pi q}{\theta ^{2}}a^{2}\exp [i(\epsilon _{2}^{a}-\epsilon
_{1}^{a}+\theta )\Omega t/\theta ],  \notag
\end{eqnarray}%
where we used formulas (\ref{Kepler}) for $\Omega $ and for $r_{b}$ on the
second line of Eq. (\ref{angularmomentum}). Notice that we chose the same $q$
in Eq. (\ref{multiscalesolution}) to cancel the fast oscillation in Eq. (\ref%
{angularmomentum}). As illustrated in Fig. 1, the fast spinning angular
momentum should rotate with the circular orbit at the frequency $\Omega $,
such that its phase should be equal to $\Omega t$. This phase-velocity
condition imposed on Eq. (\ref{angularmomentum}) yields the following
resonance condition 
\begin{equation}
\epsilon _{1}^{a}-\epsilon _{2}^{a}=0  \label{resonance1}
\end{equation}%
The calculation of $\epsilon _{1}^{a}$ and $\epsilon _{2}^{a}$ necessitates
expanding the equations of motion to quadratic order to include the
nonlinear stiff terms (the linear stability of the stiff torus). A purely
harmonic solution to these \emph{nonlinear }equations of tangent dynamics
exists because the nonlinear stiff terms introduce a correction proportional
to $a/\theta $ into Eqs. (\ref{norZ}) and (\ref{norXY}), and the complete
calculation will be given elsewhere. We expect nevertheless the
quasi-degeneracy to be preserved because the stiff limit depends basically
on the time light takes to travel between the particles. Once the stiff
torus is very close to the circular orbit, $\epsilon _{1}^{a}$ and $\epsilon
_{2}^{a}$ should differ from $\epsilon _{1}$ and $\epsilon _{2}$ by a
correction which is essentially the same because of the quasi-degeneracy
property, plus a correction of order $\theta $, such that the difference can
be well approximated by%
\begin{equation}
\epsilon _{1}-\epsilon _{2}\simeq \epsilon _{1}^{a}-\epsilon
_{2}^{a}+b\theta =b\theta   \label{approximation}
\end{equation}%
Condition (\ref{resonance1}) determines an orbital frequency proportional to
a difference of two eigenvalues, a Rydberg-Ritz-like formula%
\begin{equation}
\Omega =\Omega (\epsilon _{1}-\epsilon _{2})/b\theta =\mu \frac{\theta ^{2}}{%
b}(\epsilon _{1}-\epsilon _{2}).  \label{Rydberg}
\end{equation}%
where we have used Kepler's law of Eq. (\ref{Kepler}). The exact calculation
of $b$ demands the inclusion of the nonlinear terms into Eqs. (\ref{norZ})
and (\ref{norXY}) and shall be given elsewhere.\ Since $b$ must be of order
one, we henceforth set $b=-1$ into Eq. (\ref{approximation}) as a
qualitative approximation, yielding 
\begin{equation}
\epsilon _{1}-\epsilon _{2}+\theta =0,  \label{ressob}
\end{equation}%
which has the solutions listed in Table 1. Resonance (\ref{resonance1}) can
also be derived from the solvability of the multiscale asymptotic solution
with inclusion of the nonlinear terms by Freedholms's alternative\cite{PRA},
and shall be given elsewhere \cite{to be published}. In Ref. \cite%
{dissipaFokker} we postulated heuristically a similar type of resonance for
an analogous electromagnetic-like dynamical system, a resonance involving
the real parts of the $\lambda ^{\prime }s$. The resonant orbits of \ Ref. 
\cite{dissipaFokker} were in the atomic scale and had the same qualitative
QED-like dependence with the integer $q$, which seems to be a generic
feature of these stiff electromagnetic resonances of the two-body problem.
The root-searching problem of Eq. (\ref{ressob}) is well posed and for each
integer $q$ it turns out that one can find a pair of the form (\ref{pair})
if one sacrifices $\theta $ in Eqs. (\ref{norZ}) and (\ref{norXY}), i.e., $%
\theta $ must be quantized! According to QED, circular Bohr orbits have
maximal angular momenta and a radiative selection rule ( $\Delta l=\pm 1$)
restricts the decay from level $k+1$ to level $k$ only, i.e., circular
orbits emit the first line of each spectroscopic series (Lyman, Balmer,
Ritz-Paschen, Brackett, etc...), the third column of Table 1. We have solved
Eq. (\ref{ressob}) and Eqs. (\ref{norZ}) and (\ref{norXY}) with a Newton
method in the complex $\lambda $-plane. Every angular momentum $1/\theta $
determined by Eq. (\ref{ressob}) has the correct atomic magnitude for any $q$%
. The subset of resonant orbits listed in Table 1 has orbital frequencies
near a circular line of QED with a few percent deviation.

\begin{tabular}{|l|l|l|}
\hline
$l_{z}=\theta ^{-1}$ & $137^{3}\theta ^{2}(\epsilon _{2}-\epsilon _{1})$ & $%
w_{QED}$ \\ \hline
185.99 & 3.996$\times $10$^{-1}$ & 3.750$\times $10$^{-1}$ \\ \hline
307.63 & 8.831$\times $10$^{-1}$ & 6.944$\times $10$^{-2}$ \\ \hline
475.08 & 2.398$\times $10$^{-2}$ & 2.430$\times $10$^{-2}$ \\ \hline
577.99 & 1.331$\times $10$^{-2}$ & 1.125$\times $10$^{-2}$ \\ \hline
694.77 & 7.667$\times $10$^{-3}$ & 6.111$\times $10$^{-3}$ \\ \hline
826.22 & 4.558$\times $10$^{-3}$ & 3.685$\times $10$^{-3}$ \\ \hline
973.12 & 2.790$\times $10$^{-3}$ & 2.406$\times $10$^{-3}$ \\ \hline
1136.27 & 1.752$\times $10$^{-3}$ & 1.640$\times $10$^{-3}$ \\ \hline
1316.44 & 1.127$\times $10$^{-3}$ & 1.173$\times $10$^{-3}$ \\ \hline
1514.40 & 7.403$\times $10$^{-3}$ & 8.678$\times $10$^{-4}$ \\ \hline
1730.93 & 4.958$\times $10$^{-3}$ & 6.600$\times $10$^{-4}$ \\ \hline
1966.77 & 3.379$\times $10$^{-4}$ & 5.136$\times $10$^{-4}$ \\ \hline
2222.70 & 2.341$\times $10$^{-4}$ & 4.076$\times $10$^{-4}$ \\ \hline
\end{tabular}%
\begin{tabular}{|l|}
\hline
$q$ \\ \hline
7 \\ \hline
9 \\ \hline
11 \\ \hline
12 \\ \hline
13 \\ \hline
14 \\ \hline
15 \\ \hline
16 \\ \hline
17 \\ \hline
18 \\ \hline
19 \\ \hline
20 \\ \hline
21 \\ \hline
\end{tabular}

Caption to Table 1: Numerically calculated angular momenta $l_{z}=\theta
^{-1}$ in units of $e^{2}/c$, orbital frequencies in atomic units $%
137^{3}\theta ^{2}(\epsilon _{2}-\epsilon _{1})$, circular lines of QED in
atomic units $w_{QED}\equiv \frac{1}{2}(\frac{1}{k^{2}}-\frac{1}{(k+1)^{2}})$
, and the values of the integer $q$ of Eq. (\ref{pair}).

We identified the necessary existence of a fast timescale in the
electromagnetic two-body motion near circular orbits and explored the
consequences of balancing the associated fast dynamics. The balancing of the
fast dynamics predicts magnitudes in qualitative and quantitative agreement
with QED. The stability analysis naturally involves an integer and the
angular momenta are quantized. The emission lines are in reasonable
quantitative agreement with those of QED and the angular momentum of the
fast spinning motion is of the order of the lowest orbital angular momentum,
in agreement with QED where the electron has an angular momentum of spin
given by $\sqrt{3}\hbar /2$

\ Caption to Figure 1: Coulombian guiding-center circular orbit of each
particle, dashed lines. The trajectory of each particle is illustrated as
the fast spiral encircling each guiding-center circular orbit (solid lines).
Arbitrary units, distances are not on scale, illustrative purposes only.

We thank Savio B. Rodrigues and Reginaldo Napolitano for discussions.

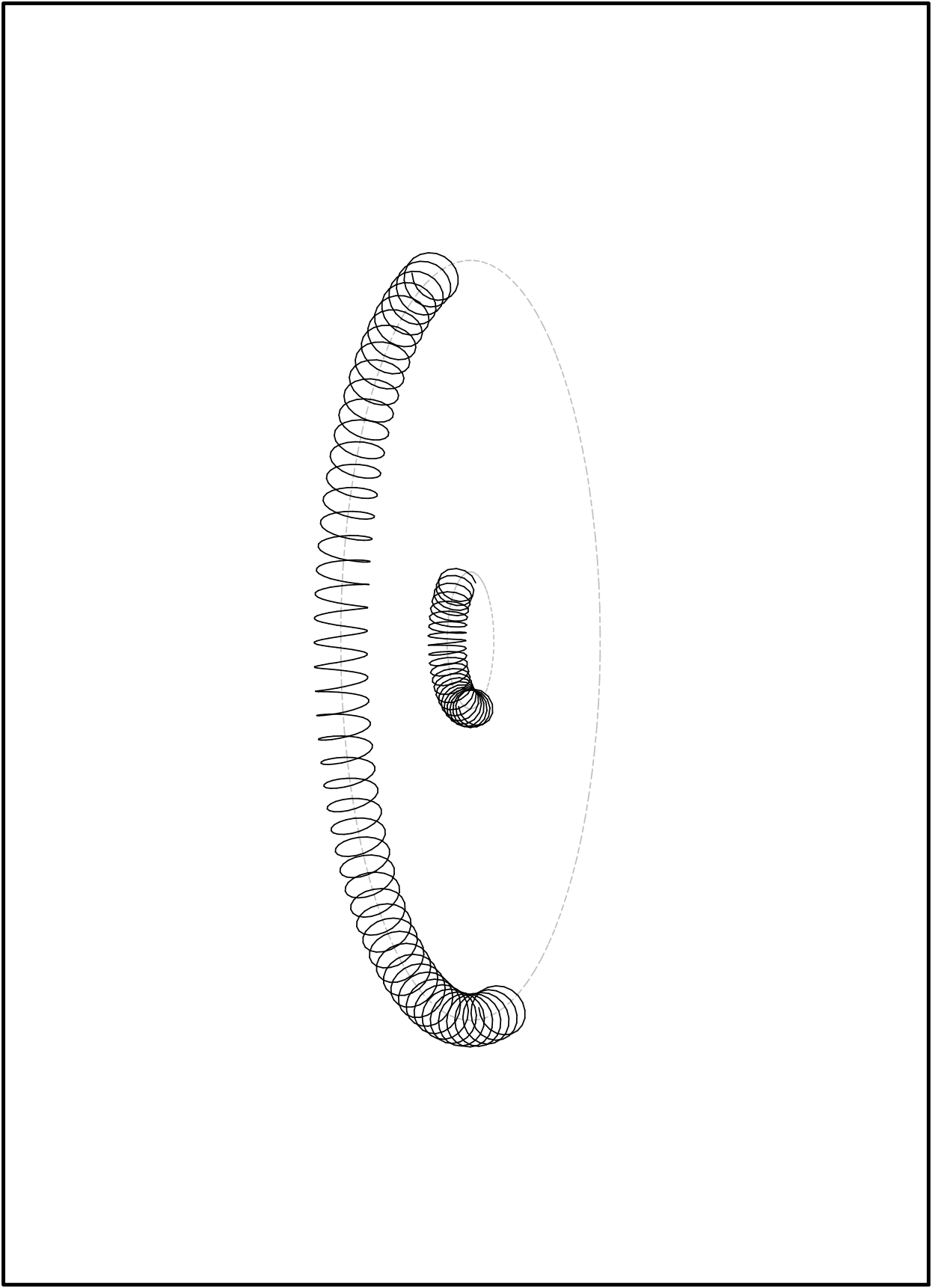
\end{document}